\documentclass{nature}
\usepackage{epsfig}
\usepackage{amsmath}

\begin{document}

\title{Experimental Demonstration of a Heralded Entanglement Source}
\author{Claudia Wagenknecht$^{1\dagger}$, Che-Ming Li$^{1,2\dagger}$, Andreas Reingruber$^{1}$,
Xiao-Hui Bao$^{1,3}$, Alexander Goebel$^{1}$, Yu-Ao Chen$^{1,3}$,
Qiang Zhang$^{1}$, Kai Chen$^{3\star}$, and Jian-Wei Pan$^{1,3\star}$}

\maketitle

\begin{affiliations}
\item
Physikalisches Institut,
Ruprecht-Karls-Universit\"{a}t Heidelberg, Philosophenweg 12,
69120 Heidelberg, Germany

\item
Department of Physics and
National Center for Theoretical Sciences, National Cheng Kung
University, Tainan 701, Taiwan

\item
Hefei National Laboratory
for Physical Sciences at Microscale and Department of Modern
Physics, University of Science and Technology of China, Hefei,
Anhui 230026, China
\\$^\dagger$These authors contributed equally to this work.
\\$^\star$e-mail: kaichen@ustc.edu.cn; jian-wei.pan@physi.uni-heidelberg.de
\end{affiliations}

\begin{abstract}
The heralded generation of entangled states is a long-standing goal
in quantum information processing, since it is indispensable for a
number of quantum protocols\cite{Nielsen00book,Bouwmeester00book}.
Polarization entangled photon pairs are usually generated through
spontaneous parametric down conversion (SPDC)\cite{Kwiat} whose
emission, however, is probabilistic. Their applications are
generally accompanied with post-selection and destructive photon
detection. Here, we report a source of entanglement generated in an
event-ready manner by conditioned detection of auxiliary
photons\cite{SliwaBanaszek}. This scheme profits from the stable and
robust properties of SPDC and requires only modest experimental
efforts. It is flexible and allows to significantly increase the
preparation efficiency by employing beam splitters with different
transmission ratios. We have achieved a fidelity better than $87 \%$
and a state preparation efficiency of $45 \%$ for the source. This
could offer promising applications in essential photonics-based
quantum information tasks, and particularly enables optical quantum
computing by reducing dramatically the computational
overhead\cite{BrowneRudolph,kok07-79}.
\end{abstract}

Quantum entanglement is one of the key resources in quantum
information and quantum foundation.
Besides its fundamental interest to reveal fascinating aspects of quantum mechanics,
they are also crucial for a variety of quantum information
tasks\cite{Nielsen00book,Bouwmeester00book}. In particular, photonic
entangled states are robust against decoherence, easy to manipulate
and show little loss, both in fiber and free-space transmission, and
thus are exceptionally well suitable for long distance quantum
communication and linear optical quantum
computing\cite{Jian-weiRMP,kok07-79}. Consequently, an event-ready
source for entangled photonic states is of great importance, both
from the fundamental and the practical point of views. Entanglement
sources based on the probabilistic generation process of SPDC allow
for demonstrations of a number of quantum protocols, but do not
permit on-demand applications, deterministic quantum computing and
significantly limit the efficiency of multi-photon experiments.
Alternative solutions, such as the controlled biexciton emission of
a single quantum dot\cite{QD1,QD2,Stevenson06} or the creation of
heralded entanglement from atomic ensembles\cite{Bo98}, face severe
experimental disadvantages, such as liquid-helium temperature
environment and large-volume setups.

There has been considerable progress towards the demonstration
of heralded photonic Bell pairs. The scheme by Knill, Laflamme and
Milburn (KLM)\cite{KLM} provides a theoretical breakthrough as proof that
efficient quantum computing is possible with linear optics.
Although the KLM scheme allows the nearly non-probabilistic creation
of entanglement, the method they use is still intrinsically probabilistic.
The fact that the KLM scheme uses a single photon source,
perfect photon-number-resolving detectors and moreover requires a
large computational capacity makes it barely accessible experimentally.
The proposal of Browne and Rudolph\cite{BrowneRudolph} comprises a significant
advance in achieving experimental implementation by using photonic
Bell pairs as the primary resource and experimentally realistic
detectors. Using their proposal, the number of optical operations
per logical two-qubit gate reduces to $\sim100$, in contrast to the original
KLM scheme, which would have $\sim100,000$ (refs 5,6,12). Central
to such a dramatic improvement is the use of a heralded entanglement
source\cite{BrowneRudolph,kok07-79}.

Various ideas based on conditional detection of auxiliary photons or
multi-photon interference were recently proposed to overcome the
probabilistic character of
SPDC\cite{KokPhD,kok00-62,SliwaBanaszek,Pittman03,Hnilo05,Eisenberg05,Walther05}.
Following this line, we demonstrate an experimental realization of a
heralded entangled photon source by adopting the proposal of
\'{S}liwa and Banaszek\cite{SliwaBanaszek}. This source provides a
substantial advance over the general methods by using linear
optics\cite{KLM,KokPhD}. In the experiment, we only use commercial
threshold single photon counting modules (SPCM) as detectors and
passive linear optics. The source is feasible to support on-demand
applications, such as the controlled storage of photonic
entanglement in quantum memory\cite{yuao} to realize the quantum
repeater scheme\cite{Briegel}. Moreover, it is suitable to serve
on-chip waveguide quantum circuit applications which promise new
technologies in quantum
optics\cite{O'BrienIII}.

To demonstrate the basic principle of the heralded entangled photon source, we illustrate the
scheme of \'{S}liwa and Banaszek\cite{SliwaBanaszek} in Fig.~1. With an input
of SPDC source emitted from modes $a,b$, the scheme herald an entangled photon pair in
$c,d$ modes conditioned by triggers of four photons in $e,f$ modes. As an input of the
optical circuit, three-pair component of the down converted photons entangled in
polarizations is utilized. The quantum state of the three-pair photon term is given
by\cite{kok00-61}
\begin{equation}
\left|\Psi_3\right\rangle=\frac{1}{12}({\hat{a}^\dag}_x{\hat{b}^\dag}_y
-{\hat{a}^\dag}_y{\hat{b}^\dag}_x)^3\left|vac\right\rangle,
\label{psi3}
\end{equation}
where $\left|vac\right\rangle$ denotes the vacuum state, and
$\hat{a}^\dag$ and $\hat{b}^\dag$ are the creation operators of
photons in the modes $a$ and $b$. Horizontal and vertical
polarization are represented by $x$ and $y$, respectively. The optical
circuit (see the Methods section) transforms $\left|\Psi_3\right\rangle$
into\cite{SliwaBanaszek}:
\begin{equation}
\left|\Psi_3'\right\rangle =
\frac{1}{\sqrt{2}}RT^{2}\left|\theta\right\rangle_\textbf{t}\left|\Phi^{+}
\right\rangle_\textbf{s}+\sqrt{1-\frac{T^{4}R^{2}}{2}}\left|\Gamma\right\rangle_{ts}.
\label{psi3'}
\end{equation}
The first term of Eq.~(\ref{psi3'}) is composed of a tensor product of two states: the state
$\left|\theta\right\rangle_\textbf{t}={\hat{e}^\dag}_x{\hat{e}^\dag}_y{\hat{f}^\dag}_{x'}
{{\hat{f}^\dag}_{y'}}\left|vac\right\rangle$ denoting one photon in each of
the four trigger modes, and the maximally entangled photon pair in the output modes
\begin{equation}
\left|\Phi^{+}\right\rangle_{\textbf{s}}=\frac{1}{
\sqrt{2}}({\hat{c}^\dag}_x{\hat{d}^\dag}_x+{\hat{c}^\dag}_y{\hat{d}^\dag}_y)
\left|vac\right\rangle.
\end{equation}
The normalized state $\left|\Gamma\right\rangle_{ts}$ is a
superposition of all states that do not exactly have one photon in
each of the trigger mode $\hat{e}_{x}$, $\hat{e}_{y}$,
$\hat{f}_{x'}$ and $\hat{f}_{y'}$. Hence, the scheme for heralded
entanglement source is clearly based on the fact that when detecting
a coincidence of four single photons in the trigger modes
$({\hat{e}}_x,{\hat{e}}_y;{\hat{f}}_{x'}, {{\hat{f}}_{y'}})$, the
two photons in the output modes $(\hat{c}_x,\hat{c}_y)$ and
$(\hat{d}_x,\hat{d}_y)$ nondestructively collapse to the maximally
entangled state $\left|\Phi^{+}\right\rangle_{\textbf{s}}$.

In the experiment, we will generate the three-pair photon states
(\ref{psi3}) by a photon source (see the Methods section) and
consequently implement the transformation of the linear optical
circuit. A schematic diagram of our experimental set-up is shown in
Fig.~2, which is based on the proposal of Ref.~4.

When taking all of experimental imperfections into account (see the Methods section), it is crucial to evaluate the
performance this source. Therefore, we have measured the state preparation efficiency and its fidelity, where the
efficiency is defined by the number of heralded photon pairs created from the source per trigger signal. For an
ideal case, one trigger signal of a fourfold single-photon coincidence perfectly heralds one photon pair creation.
In our experiment performed with standard SPCMs, obviously, additional terms yielding triggers will thus result in
a reduction of the preparation efficiency. To overcome this obstacle, we limit their emergence by decreasing the
transmission coefficients of the beamsplitter. In this regime, the probability of transmitting more than the
minimum number of photons to the trigger becomes lower and as such, the danger of under counting photons in the
trigger detectors decreases. However, enhancing the preparation efficiency in this way will lower the over all
preparation rate.

In order to show the relation between the efficiency of state preparation and the transmission
coefficients of the partial reflecting beam splitters, we have chosen BS with three
different reflection/transmission ($R/T$) ratios: $48.6/51.4$, $57.0/43.0$ and
$68.5/31.5$ in the experiment (Fig.~2). In what follows we will denote them by
50/50, 60/40 and 70/30, respectively, for short. This relation is shown in Fig.~3.
The experimental efficiency can be straightforwardly represented as
the following relation by the number of triggers $n_{t}$, the average detection efficiency
$\eta_{s}$ for output states, the number of six-fold coincidences $n_{s}$ among four trigger modes
and two output modes
\begin{equation}
\text{eff}_{exp}=\frac{n_{s}}{n_{t}\eta_{s}^{2}}.
\label{effexp}
\end{equation}
For each experimental detection efficiencies $\eta_{s}$ and $R$/$T$ ratio: $0.129$ (50/50),
$0.133$ (60/40) and $0.15$ (70/30), the average coincidence counts $(n_{s},n_{t})$
observed per $10$ hours are: $(37,9710)$, $(37,4940)$ and $(14,1347)$, respectively. As can
be seen from Fig.~3, the experimental results are highly consistent
with theoretical estimation (see Supplementary Information):
\begin{equation}
\text{eff}_{theory}=\frac{R^2}{(1- \eta_{t} T/2)^2}, \label{effth}
\end{equation}
where  $\eta_{t}$ represents the average detection efficiency for
trigger photons. Thus, with this setup we have significantly
improved the preparation efficiency in comparison with the one
provided by the standard procedure through SPDC. One can consider single input pulses of UV
laser and output photon pairs of SPDC as trigger signals and output
states, respectively. The probability of generating one
entangled photon pair per UV pulse means the preparation
efficiency of the standard procedure through SPDC\cite{kok00-61}.

To quantify the entanglement of the output photons and evaluate how
the prepared state is similar to the state
$\left|\Phi^{+}\right\rangle\!_{\textbf{s}}$, we have determined the
state fidelity by analyzing the polarization state of the photons in
the modes $(\hat{c},\hat{d})$ in the three complementary bases:
linear ($H/V$), diagonal ($+/-$), and circular ($R/L$). For an
experimental state $\hat{\rho}$, the fidelity is explicitly defined by
\begin{eqnarray}
F & = & \text{Tr}(\hat{\rho}\left|\Phi^{+}\right\rangle\!_{\textbf{ss}}\!\left\langle
\Phi^{+}\right|)\nonumber\\
  & = & \frac{1}{4}(1+\langle \hat{\sigma}_x\hat{\sigma}_x\rangle-
  \langle \hat{\sigma}_y\hat{\sigma}_y\rangle+\langle \hat{\sigma}_z\hat{\sigma}_z\rangle),
\label{fidelity}
\end{eqnarray}
where $\left|\Phi^{+}\right\rangle \!_{\textbf{ss}}\!\left\langle \Phi^{+}\right|=
\frac{1}{4}(\hat{I}+\hat{\sigma}_x\hat{\sigma}_x-\hat{\sigma}_y\hat{\sigma}_y
+\hat{\sigma}_z\hat{\sigma}_z)$, ${\hat\sigma}_z=\left|H\right\rangle\!\!\left\langle H
\right|-\left|V\right\rangle\!\!\left\langle V\right|$,
${\hat\sigma}_x=\left|+\right\rangle\!\!\left\langle+\right|-\left|-
\right\rangle\!\!\left\langle -\right|$ and
${\hat\sigma}_y=\left|R\right\rangle\!\!\left\langle R\right|-
\left|L\right\rangle\!\!\left\langle L\right|$.
Eq.~(\ref{fidelity}) implies that we can obtain the fidelity of the
prepared state $\hat{\rho}$ by consecutively carrying out three
local measurements ${\hat\sigma}_x {\hat\sigma}_x$, ${\hat\sigma}_y
{\hat\sigma}_y$ and ${\hat\sigma}_z {\hat\sigma}_z$ on the photons
in the output modes $(\hat{c}_x,\hat{c}_y)$ and
$(\hat{d}_x,\hat{d}_y)$ (see the Methods section). In the experiment, we only used threshold
SPCMs to perform measurements. The experimental results
are shown in Fig.~4. The experimental integration time for each
local measurement, with respect to different reflection/transmission
ratio of the BS, took about: $19$ h (50/50), $17$ h (60/30) and $36$
h (70/30). For all three splitting ratios, we recorded more than
about 50 events of desired six-photon coincidences for each local
measurement: $\sim65$ (50/50), $\sim58$ (60/30) and $\sim62$
(70/30). As can be seen from Table
1, the measured values for the fidelity are sufficient to violate
CHSH-type Bell's inequality\cite{CHSH1969} for Werner states by
three standard deviations. Since we only used threshold SPCMs as
detectors, the measured coincidences are then affected by unwanted
events. In our experiment, the effect of the dark count rate in the
detectors on the six-fold coincidence is rather
small. (About the dark count contribution,
the main part is that one detector is triggered by dark counts, and
the other five detectors are triggered by the down conversion
photons. Given a three-pair state, the probability of generating a
six-fold coincidence count within any particular coincidence window
is about $S\sim \eta^{6}$, whereas the leading dark count
contribution is about $S_{d}\sim \eta^{5}D$, where $D=n_{d}t$,
$n_{d}$ is the average dark count rate of detector, and $t$ denotes
the coincidence window. In our experiment, we have $n_{d}\sim 300$
Hz and $t=12\times10^{-9}$ sec. Then it is clear that the dark count
rate in detectors contribute a very small part of the six-fold
coincidences: $S_{d}/S=n_{d}t/\eta\sim 2\times10^{-5}$. Here
$\eta=15\%$ is used for the estimation.)

In conclusion, we have demonstrated a heralded source for
photonic entangled states, which is capable of circumventing the
problematic issue of probabilistic nature of SPDC. Such source is
based on the well known technique of type-II SPDC, which is robust,
stable and needs only modest experimental efforts by using standard
technical devices. Photon number resolving detectors are not
involved in the setup, and therefore we do not endure the
restriction inherent to other schemes for implementing heralded
entanglement sources \cite{KokPhD,Pittman03}. To evaluate the
performance of our source, we have measured the fidelity of the
output state, and demonstrated the relation between the amplitude
reflection coefficient of the used beam splitters and the
preparation efficiency of the source. A fidelity better than $87 \%$
and a state preparation efficiency of $45 \%$ are achieved. For future
applications, the simple optical circuit of our source could be
miniaturized by an integrated optics architecture on a chip using
the silica-on-silicon technique\cite{O'BrienI}. Using waveguides
instead of bulk optics would be beneficial to stability, performance
and scalability\cite{O'BrienII,O'BrienIII}.
We note that during the preparation of the manuscript presented here, we
learned of a parallel experiment by Barz {\it et al.} \cite{Walther2010}.

\section*{Methods}

\noindent \textbf{Optical circuit.} The
transformation of the optical circuit consists of BS and HWP
operations. The BS operation describes the following transformation
of the annihilation operators of the modes $\hat{a}_{k}$ and
$\hat{b}_{k}$ (note that we use annihilation operators to denote the
corresponding modes):
$\hat{a}_{k}=\sqrt{R}\hat{c}_{k}+\sqrt{T}\hat{e}_{k}$ and
$\hat{b}_{k}=\sqrt{R}\hat{d}_{k}+\sqrt{T}\hat{f}_{k}$, for $k=x,y$.
$R$ ($T$) is the amplitude reflection (transmission) coefficient of
the BS. For the modes $\hat{f}_{x}$ and $\hat{f}_{y}$, the
transformation of HWP at $-22.5^{\circ}$ is defined by:
$\hat{f}_{x}=(\hat{f}_{x'}-\hat{f}_{y'})/\sqrt{2}$ and
$\hat{f}_{y}=(\hat{f}_{x'}+\hat{f}_{y'})/\sqrt{2}$. The optical circuit is able to prevent false signals rising from
two-pair emission. This is an important feature of the scheme\cite{SliwaBanaszek} since
the creation probabilities for two pairs are much larger than for
three pairs. Furthermore, contributions from the higher order terms
of SPDC can be limited by controlling the corresponding creation
probabilities\cite{SliwaBanaszek}. It is also worth noting that for
a given three-pair photon state, the probability of creating a
heralded entangled state, i.e., $T^{4}R^{2}/2$, is controllable by
changing the transmission coefficients of the BS, which can be up to
$\sim0.011$\cite{SliwaBanaszek}.

\noindent \textbf{Photon source.} The required photon pairs are generated by type-II SPDC from a
pulsed laser in a $\beta$-Barium-Borate (BBO) crystal. Here, we use
a pulsed high-intensity ultraviolet (UV) laser with a central
wavelength of $390$ nm, a pulse duration of $180$ fs and repetition
rate of $76$ MHz. For an average power of $880$ mW UV light and
after improvements in collection efficiency and stability of the
photon sources, we observe $\sim80\times10^{3}$ photon pairs per
second with a visibility of $\mathcal{V}=(91\pm 3) \%$
measured in the diagonal ($+/-$) basis.
(The visibility is defined by
$\mathcal{V}=(N_{d}-N_{ud})/(N_{d}+N_{ud})$, were $N_{d}$ ($N_{ud}$)
denotes the number of two-fold desired (undesired) coincidence
counts. Then there exists a direct connection between visibility and
fidelity of a measured state $\hat{\rho}$:
$F=\text{Tr}(\hat{\rho}\left|\Psi^{-}\right\rangle\!\!\left\langle
\Psi^{-}\right|)=\frac{1}{4}(1+\mathcal{V}_{x}+\mathcal{V}_{y}+\mathcal{V}_{z})$,
where $\mathcal{V}_{k}$ for $k=x,y,z$ denotes the visibility of
photon pair in the diagonal, circular, and linear bases,
respectively. Here $\left|\Psi^{-}\right\rangle$ is the singlet Bell
state.) Then the probability of
creating three photon pairs is about $5.7\times10^{-5}$ per pulse,
which is $\sim33$ times larger than that of the next leading order
term. The estimation of the three-pair creation probability per
pulse is based on the experimental pair generation rate and the
theoretical $n$-pair creation probability\cite{kok00-61}
$p_{n}=(n+1)\tanh^{2n}r/\cosh^{4}r$, where $r$ is a real-valued
coupling coefficient. From the two-fold coincidence measurement
result, the experimental pair generation rate is
$p'=(80\times10^{3})/(0.15^{2}\times76\times10^{6})\approx4.7\%$. We
assume that $p_{1}=p'$, $r$ can directly be derived from $p_{1}$.
Thus the estimated creation probability $p_{3}$ and $p_{4}$ are
obtained.

\noindent \textbf{Experimental imperfections.} With single photon resolving detectors and $100 \%$ detection
efficiency, one can see that the three-pair state can provide a maximally entangled photon pair in the output modes
deterministically with a 100\% probability, if and only if the remaining photons give rise to a fourfold
coincidence among the four trigger modes. With the widely used standard SPCM, one cannot discriminate pure single
photons from multi-photons which in reality leads to a significant problem of under counting photons. Accordingly,
the trigger detectors can herald a successful event even though more than two photons from either mode
$(\hat{a}_x,\hat{a}_y)$ or $(\hat{b}_x,\hat{b}_y)$ or both have been transmitted to the trigger channels.
Furthermore, experimentally we were only able to obtain an average detection efficiency of about $\eta=15\%$
resulting from limited collection and detector efficiencies. Here the mean detection efficiency is averaged over
the coupling efficiency of eight fibre couplers and the quantum efficiency of the detectors. In addition to the
imperfect detections, there are two other factors that affect the performance of our source: the non-ideal quality
of the initially prepared pairs and the higher-order terms of down-converted photons. For perfectly created pairs,
destructive two-photon interference effects\cite{Shih-Alley,Hong-Ou-Mandel,LamasLinares} will extinguish the
contribution of two-pair emission to the trigger signal. With an experimental visibility of $(91\pm 3) \%$
imperfectly created states may still give rise to a contribution of two-pair events that leads to the detection of
the auxiliary triggers. In addition, four-pair emission can again contribute to both the triggers and the output.
Although the experimentally estimated creation probability for a four-pair emission is only $\sim1.7\times 10^{-6}$
per pulse and is much smaller than the probability for a three-pair photon state $\sim5.7\times 10^{-5}$ per pulse,
four-pair contribution can lead to an error of the theoretical estimation of the expected preparation efficiency of
about $4.5 \%$. The four-pair contribution is evaluated in the same way as the three-pair state, where the limited
detection efficiency of the trigger detectors is considered in the calculation (see Supplementary Information). In
Fig.~3, the fluctuations of our experimental data mainly result from the intrinsic statistics of detector counts,
and the stability of optical alignment.

\noindent \textbf{Experimental fidelity $F$.} Every expectation value for a correlation function is
obtained by making a local measurement along a specific polarization
basis and computing the probability over all the possible events.
For instance, to get the expectation value of $RR$ correlation
$\text{Tr}(\hat{\rho}\left|RR\right\rangle\!\!\left\langle
RR\right|)$, we perform measurements along the circular basis and
then get the result by the number of coincidence counts of $RR$ over
the sum of all coincidence counts of $RR$, $RL$, $LR$ and $LL$. All
the other correlation settings are performed in the same way. The
fidelity $F$ can then directly be evaluated.

\begin{addendum}

\item [Acknowledgement]

This work was supported by the European Commission through the European Research
Council (ERC) Grant and the Specific Targeted Research Projects (STREP) project Hybrid
Information Processing (HIP), the Chinese Academy of Sciences, the National
Fundamental Research Program of China under grant no. 2006CB921900, and the National
Natural Science Foundation of China. C.W. was additionally supported by the Schlieben-
Lange Program of the ESF. The authors are grateful to Dr Xian-Min Jin for help in
improving the figures.

\item [Author Contributions]
C.W., X.-H.B., Y.-A.C., Q.Z., and J.-W.P. designed the experiment.
C.W., C.-M.L., A.R., A.G., Y.-A.C., and K.C. performed the experiment.
C.W., C.-M.L., A.R., X.-H.B., K.C., and J.-W.P. analyzed the data. C.W., C.-M.L.,
K.C. and J.-W.P. wrote the paper.

\item [Competing Interests]
The authors declare that they have no competing financial interests.

\item [Correspondence]
Correspondence and requests for materials should be addressed to K.C. and J.W.P.
\end{addendum}

\clearpage

\begin{table}
\begin{center}
\epsfig{file=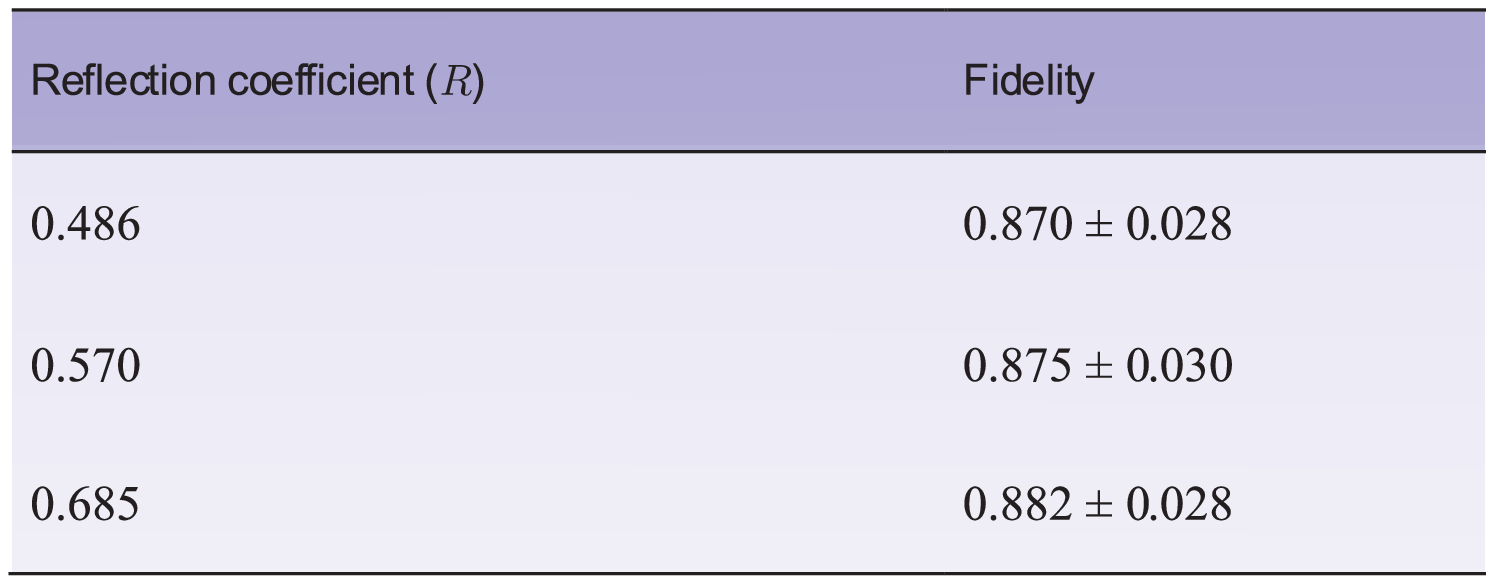,width=16cm}
\end{center}
\caption{Experimental fidelity of the entangled output state with respect to
the reflection coefficients $R$ of the beam splitters.}
\label{table1}
\end{table}

\newpage
\bigskip
\textbf{Figure 1 Schematic setup.} The heralded generation of
entangled photon pairs is implemented with the optical circuit
composed of non-polarizing partial reflecting beam splitters (BS), a
half wave plate (HWP) and two polarizing beam splitters (PBS). The
BS split mode $\hat{a}$ $(\hat{b})$ into a trigger mode $\hat{e}$
$(\hat{f})$ and an output mode $\hat{c}$ $(\hat{d})$. The auxiliary
trigger photons are detected in $(\hat{f}_{x'},\hat{f}_{y'})$ in the
diagonal ($+/-$) basis and in $(\hat{e}_x,\hat{e}_y)$ in the linear
($H/V$) basis. The setup will output an entangled photon pair after
successful triggering of the four auxiliary photons.

\bigskip

\textbf{Figure 2 Experimental setup for event-ready entanglement source.}
After emission, the longitudinal and spatial walk-off of the photons in mode $\hat{a}$ and $\hat{b}$ will be
compensated by a HWP and a correction BBO (C BBO) before the photons are directed onto the partial reflecting
beamsplitter (PRBS). To control the additional phase introduced by the PRBS we used a combination of two
quarter-wave plates (QWP) and one HWP. All photons are filtered by narrow bandwidth filters ($\Delta \lambda
\approx 3.2$ nm) and are monitored by silicon avalanche single-photon detectors. Coincidences are recorded by a
laser clocked FPGA (Field Programmable Gate Array) based coincidence unit.

\bigskip

\textbf{Figure 3 Efficiency of state preparation.} Theoretical and
experimental values of preparation efficiency for the amplitude
reflection coefficients $R=0.486, 0.570$ and $0.685$ are depicted.
The error bars are according to Poissonian statistics of counts. The
curve is a function graph of Eq.~\ref{effth} with an average
detection efficiency $\eta_{t}=0.1823$ for triggers.
$\text{eff}_{theory}$ is an increasing function of $R$ and up to
$100\%$. The quantum efficiency of detectors $q$ used is about
$60\%$. For each 50/50, 60/40, and 70/30 BS ratio, our experimental
coupling efficiencies of trigger ($p_{t}$) and signal detectors
($p_{s}$) are as follows, $(p_{t},p_{s})$: $(27.8\%,21.5\%)$,
$(28.8\%,22.2\%)$, and $(34.5\%, 25.0\%)$, respectively. Note that
$pq$ is defined as the detection efficiency $\eta$.

\bigskip

\textbf{Figure 4 Experimental data for fidelity measurements.} We
have performed a complete 3-setting local measurements for
${\hat\sigma}_z {\hat\sigma}_z$, ${\hat\sigma}_x {\hat\sigma}_x$ and
${\hat\sigma}_y {\hat\sigma}_y$, which corresponding to three
complementary bases of $|H\rangle /|V\rangle$, $|+\rangle
/|-\rangle$ and $|R\rangle /|L\rangle$, with
$|+\rangle=(|H\rangle+|V\rangle)/\sqrt{2}$,
$|-\rangle=(|H\rangle-|V\rangle)/\sqrt{2}$,
$|R\rangle=(|H\rangle+i|V\rangle)/\sqrt{2}$ and
$|L\rangle=(|H\rangle-i|V\rangle)/\sqrt{2}$.
The plots are for three different splitting ratios R/T
of the partial reflecting beamsplitters 50/50
(\textbf{a}), 60/40 (\textbf{b}) and 70/30 (\textbf{c}).
The error bars relate to Poissonian statistics of counts.

\newpage
\begin{figure}
\begin{center}
\epsfig{file=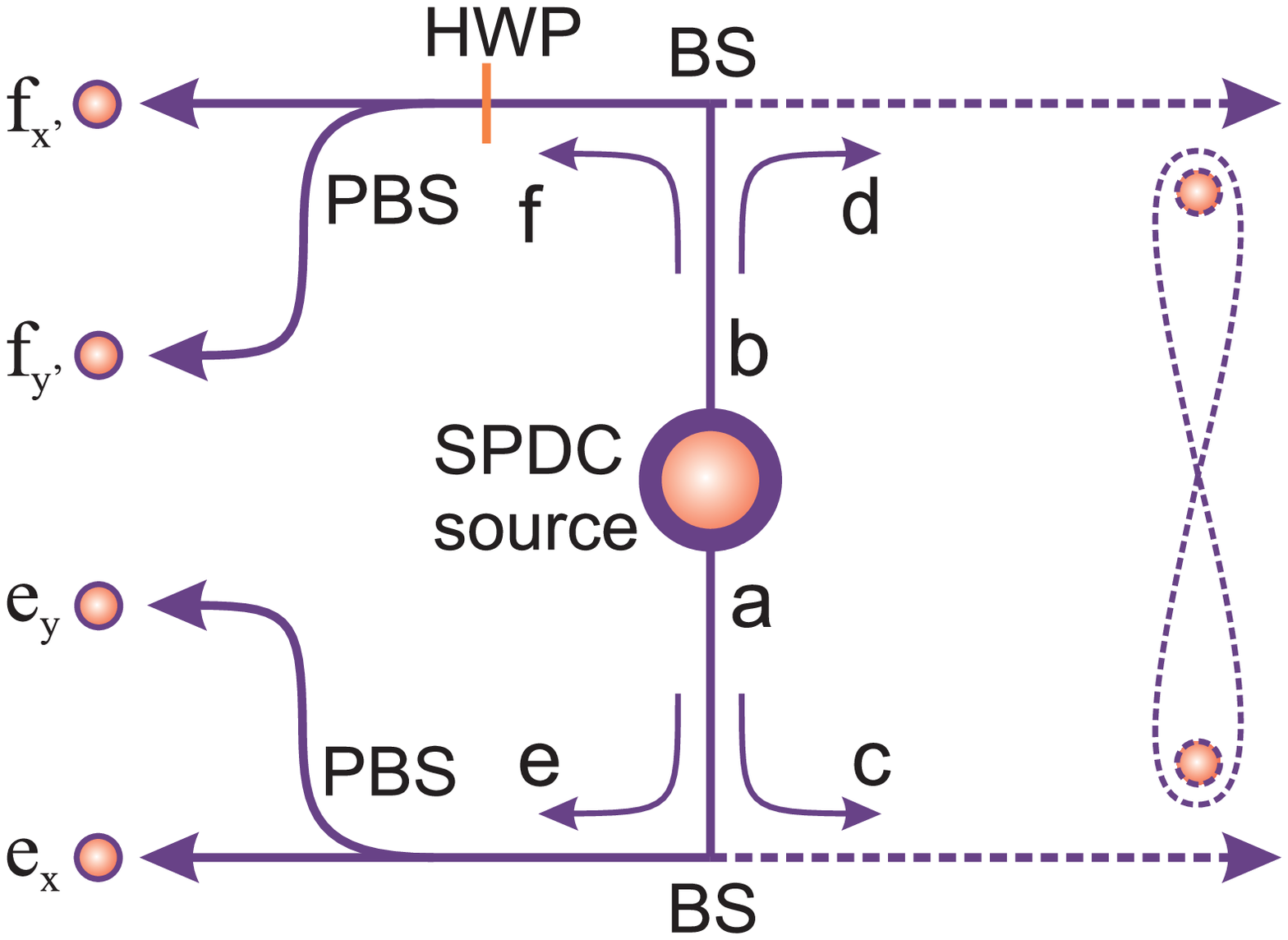,width=14cm}
\end{center}
\label{fig1-scheme}
\end{figure}

\begin{figure}
\begin{center}
\epsfig{file=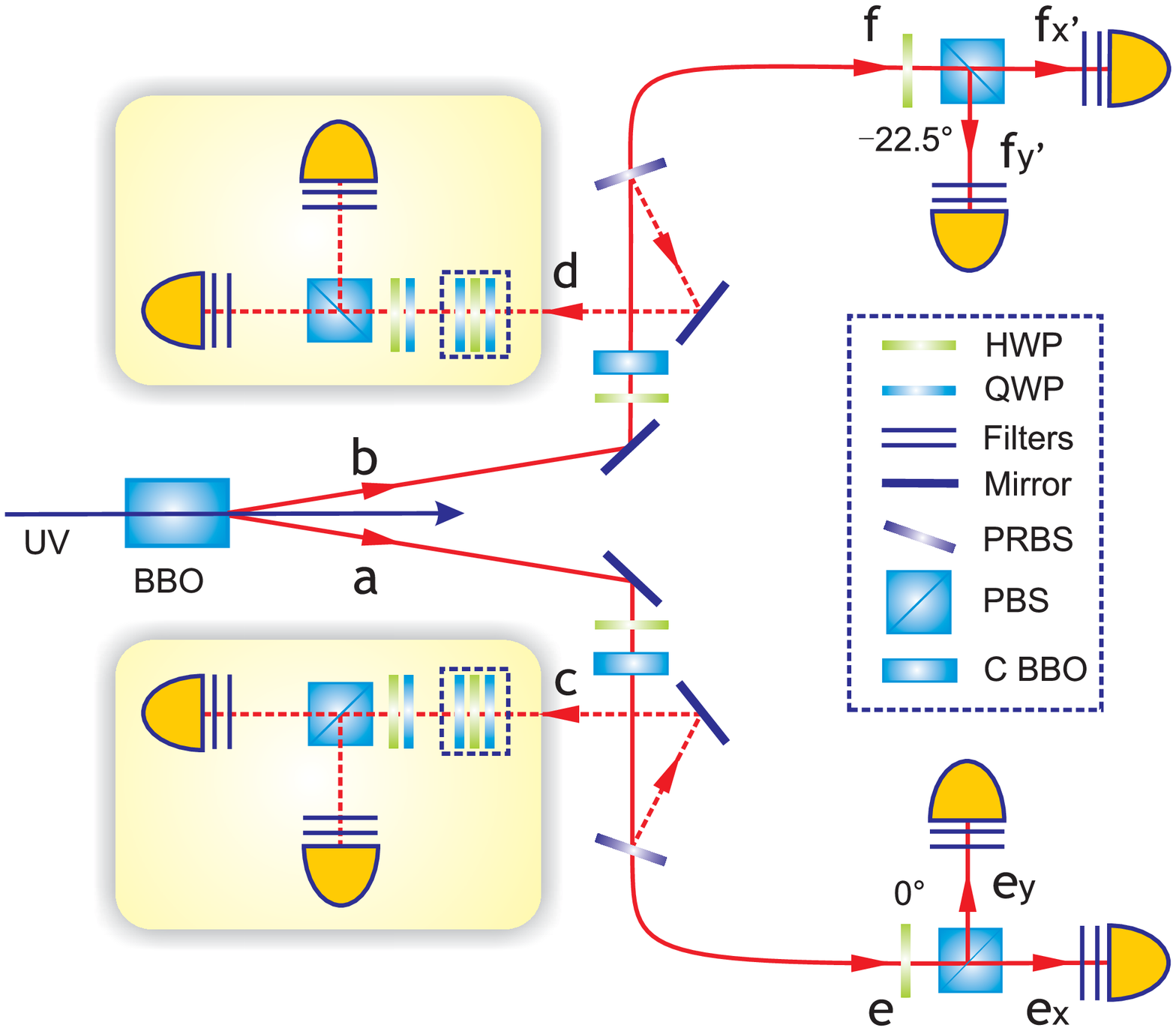,width=14cm}
\end{center}
\label{fig2-setup}
\end{figure}

\begin{figure}
\begin{center}
\epsfig{file=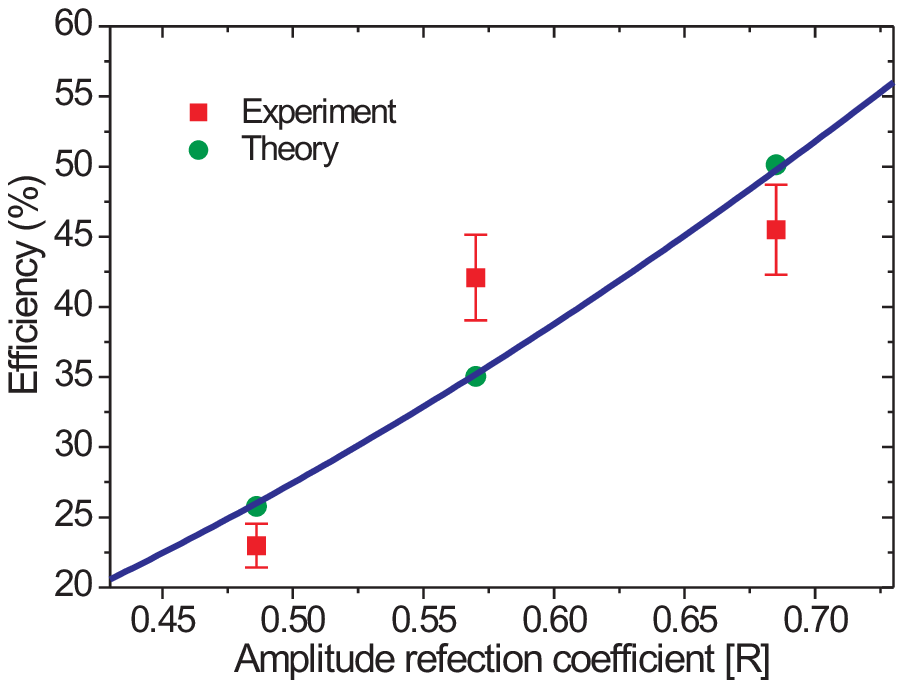,width=14cm}
\end{center}
\label{fig3-finalploteff}
\end{figure}

\begin{figure}
\begin{center}
\epsfig{file=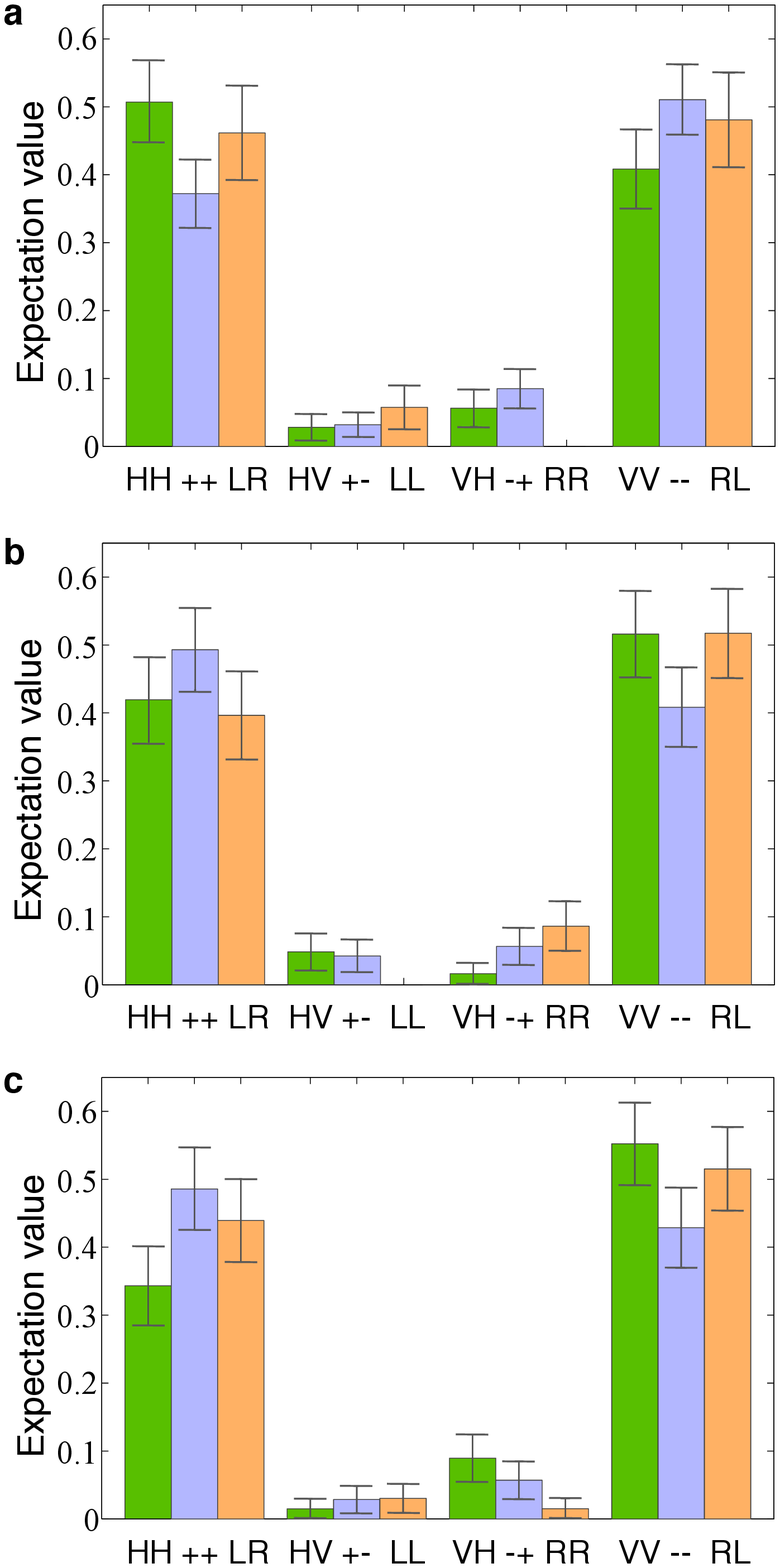,width=12cm}
\end{center}
\label{fig4-barcharts}
\end{figure}
\clearpage
\bigskip

\newpage
\bigskip

\noindent\textbf{\Large{Experimental Demonstration of a Heralded Entanglement Source\\
Supplementary Information}}

\noindent \textbf{Efficiency of state preparation ${\text{eff}}_{theory}$}

In order to
show a clear picture of the theoretical estimation, let us consider first the case
that we have ideal detection efficiency of $100\%$. Eq.~(5) will
be naturally derived afterwards. It is instructive to start with the output state
$\left|\Psi_3'\right\rangle$ in the following complete form:
\begin{equation}
\left|\Psi_{3}'\right\rangle =
\alpha\left|\theta\right\rangle_\textbf{t}\left|\Phi^{+}\right\rangle_\textbf{s} + \beta
\left|\vartheta \right\rangle_{\textbf{t}s} + \gamma  \left|\varphi \right\rangle_{t
\textbf{s}}. \tag{S1}
\end{equation}

For the first term of Eq.~(S1), as already indicated in
Eq.~(2), the appearance of the perfect four-photon trigger
state $\left|\theta\right\rangle_\textbf{t}$ will herald a photon
pair in state $\left|\Phi^{+}\right\rangle_{\textbf{s}}$ in the
output with a probability of $100 \%$. The second term
$\beta\left|\vartheta \right\rangle_{\textbf{t}s}$ represents all
additional states, which actually yield a trigger signal in all of
the $(\hat{e}_x,\hat{e}_y)$ and $(\hat{f}_{x'}\hat{f}_{y'})$ modes
generated by more than four photons, but the state in the output
$(\hat{c}_x,\hat{c}_y)$ and $(\hat{d}_x,\hat{d}_y)$ contains only a
single photon or vacuum. For completeness, the last term $\gamma
|\varphi \rangle_{t \textbf{s}}$ represents states, which do not
contribute to the trigger, but contain more than two photons in the
output modes. These states do not affect our experimental results
since without trigger signal their contribution is not recorded. All
normalization factors are summarized in $\alpha$, $\beta$ and
$\gamma$. Therefore, according to the definition of the efficiency
of state preparation, we obtain the state preparation efficiency for
perfect detections:
${\text{eff}}_{theory}=\alpha^{2}/(\alpha^{2}+\beta^{2})=R^{2}/(1-T/2)^{2}$.
Now we proceed to discuss the effect of imperfect detection. First,
we introduce the detection loss of coupling by replacing the
creation operators in the trigger modes $\hat{\textbf{t}}^{\dag}$
for $\textbf{t}=e_{x},e_{y},f_{x'},f_{y'}$ with $\sqrt{p}\
\hat{\textbf{t}}^{\dag}+\sqrt{1-p}\
\hat{\tilde{\textbf{t}}}^{\dag}$, where $p$ denotes the coupling
efficiency for the trigger photons\cite{SliwaBanaszek}. The
operators $\hat{\tilde{\textbf{t}}}^{\dag}$ describe photons that
escape from detections. Then we consider all the components of the
state $\left|\Psi_{3}'\right\rangle$ that can be collected into
trigger detectors, e.g.,
$\beta'p^{2}\sqrt{1-p}\hat{d}_{x}^{\dag}\hat{e}_{x}^{\dag}\hat{\tilde{e}}_{x}^{\dag}
\hat{e}_{y}^{\dag}\hat{f}_{x'}^{\dag}\hat{f}_{y'}^{\dag}\left|vac\right\rangle$.
Meanwhile, we take the efficiency of detector into account. For each
considered term, we therefore can obtain the corresponding
probability of giving a herald signal from its probability amplitude
and state vector. For example, for the term illustrated above the
probability is $\beta'^{2}q^{4}p^{4}(1-p)$, where $q$ denotes the
efficiency of trigger detector. Finally, by the probability of
measuring a heralded photon pair over the total probability of
generating a trigger signal, we have $\text{eff}_{theory}=R^2/(1-
pqT/2)^2$. Here $pq$ is defined as the detection efficiency for
trigger photons $\eta_{t}$. For each 50/50, 60/40 and 70/30 BS
ratios, our experimental detection efficiencies achieve: $0.167$,
$0.173$ and $0.207$, respectively.


\begin{thebibliography}{99}
\bibitem{Nielsen00book} Nielsen, M.A. \& Chuang, I. L. \textit{Quantum Computation and
Quantum Information.} (Cambridge University Press, Cambridge,
2000).

\bibitem{Bouwmeester00book}
Bouwmeester, D., Ekert, A. K. \& Zeilinger, A. \textit{The Physics
of Quantum Information.} (Springer, Berlin, 2000).

\bibitem{Kwiat}
Kwiat, P. G., Mattle, K., Weinfurter, H., Zeilinger, A., Alexander, V. S. \& Shih, Y. H. New
high-intensity source of polarization-entangled photon pairs.
\textit{Phys. Rev. Lett.} \textbf{75}, 4337-4341 (1995).

\bibitem{SliwaBanaszek}
\'{S}liwa, C. \& Banaszek, K. Conditional preparation of maximal
polarization entanglement. \textit{Phys. Rev. A} \textbf{67},
030101(R) (2003).

\bibitem{BrowneRudolph}
Browne, D. E. \& Rudolph, T. Resource-efficient linear optical
quantum computation. \textit{Phys. Rev. Lett.} \textbf{95}, 010501
(2005).

\bibitem{kok07-79}
Kok, P. {\it et al.} Linear optical quantum computing with
photonic qubits. \textit{Rev. Mod. Phys.} \textbf{79}, 135-174
(2007).

\bibitem{Jian-weiRMP}
Pan, J.-W., Chen, Z.-B., Zukowski, M., Weinfurter, H. \&
Zeilinger, A. Multi-photon entanglement and interferometry.
\textit{Rev. Mod. Phys.}
Preprint at http://arXiv:0805.2853 (2008).

\bibitem{QD1}
Benson, O., Santori, C., Pelton, M. \& Yamamoto, Y.
Regulated and Entangled Photons from a Single Quantum Dot.
\textit{Phys. Rev. Lett.} \textbf{84}, 2513-2516 (2000).

\bibitem{QD2}
Akopian, N. {\it et al.} Entangled Photon Pairs from Semiconductor Quantum Dots.
\textit{Phys. Rev. Lett.} \textbf{96}, 130501 (2006).

\bibitem{Stevenson06}
Stevenson, R. M. {\it et al.} A semiconductor source of triggered
entangled photon pairs. \textit{Nature} \textbf{439}, 179-182 (2006).

\bibitem{Bo98}
Zhao, B., Chen, Z.-B., Chen, Y.-A., Schmiedmayer, J. \& Pan, J.-W.
Robust creation of entanglement between remote memory qubits.
\textit{Phys. Rev. Lett.} \textbf{98}, 240502 (2007).

\bibitem{KLM}
Knill, E., Laflamme, R. \&  Milburn, G. J.  A scheme for efficient
quantum computation with linear optics. \textit{Nature}
\textbf{409}, 46-52 (2001).

\bibitem{KokPhD}
Kok, P. \textit{State Preparation in Quantum Optics.}, PhD thesis,
University of Wales (2000).

\bibitem{kok00-62}
Kok, P. \& Braunstein, S. Limitations on the creation of maximal
entanglement. \textit{Phys. Rev. A} \textbf{62}, 064301 (2000).

\bibitem{Pittman03}
Pittman, T.B. {\it et al.} Heralded two-photon entanglement from
probabilistic quantum logic operations on multiple parametric
down-conversion sources. \textit{IEEE J. Sel. Top. Quantum
Electron.} \textbf{9}, 1478-1481 (2003).

\bibitem{Hnilo05}
Hnilo, A.A. Three-photon frequency down-conversion as an
event-ready source of entangled states. \textit{Phys. Rev. A}
\textbf{71}, 033820 (2005).

\bibitem{Eisenberg05}
Eisenberg, H.S., Hodelin, J.F., Khoury, G. \& Bouwmeester, D.
Multiphoton path entanglement by nonlocal bunching. \textit{Phys.
Rev. Lett.} \textbf{94}, 090502 (2005).

\bibitem{Walther05}
Walther, P., Aspelmeyer M. \& Zeilinger, A. Heralded generation of
multiphoton entanglement. \textit{Phys. Rev. A} \textbf{75},
012313 (2007).

\bibitem{yuao}
Chen, Y.-A. {\it et al.}  J.  Memory-built-in quantum
teleportation with photonic and atomic qubits. \textit{Nature
Physics} \textbf{4}, 103-107 (2008).

\bibitem{Briegel}
Briegel, H.-J., D\"{u}r, W., Cirac, J. I., \& Zoller, P. Q.
Quantum repeaters: the role of imperfect local operations in
quantum communication. \textit{Phys. Rev. Lett.} \textbf{81}, 5932-5935
(1998).

\bibitem{O'BrienIII}
Matthews, J.C.F., Politi, A., Stefanov, A., \& O'Brien, J. L.
Manipulation of multiphoton entanglement in waveguide quantum
circuits. \textit{Nature Photonics} \textbf{3}, 346-350 (2009).

\bibitem{kok00-61}
Kok, P. \& Braunstein, S. Postselected versus nonpostselected
quantum teleportation using parametric down-conversion.
\textit{Phys. Rev. A} \textbf{61}, 042304 (2000).

\bibitem{CHSH1969} Clauser J., Horne, M., Shimony, A. \& Holt, R.
Proposed experiment to test local hidden-variable theories.
\textit{Phys. Rev. Lett.} \textbf{23}, 880-884 (1969).


\bibitem{O'BrienI}
Politi, A., Cryan, M. J., Rarity, J. G., Yu, S. \& O'Brien,J. L.
Silica-on-silicon waveguide quantum circuits. \textit{Science}
\textbf{320}, 646-649 (2008).

\bibitem{O'BrienII}
Clark, A. S., Fulconis, J., Rarity, J. G., Wadsworth, W. J. \&
O'Brien, J. L. All-optical-fiber polarization-based quantum logic
gate. \textit{Phys. Rev. A} \textbf{79}, 030303(R) (2009).

\bibitem{Walther2010}
Barz, S., Cronenberg, G., Zeilinger, A. \& Walther, P.
Heralded generation of entangled photon pairs. \textit{Nature Photon.} (in the press).

\bibitem{Shih-Alley}
Shih, Y.H. \& Alley, C.O. New type of Einstein-Podolsky-Rosen-Bohm
experiment using pairs of light quanta produced by optical
parametric down conversion. \textit{Phys. Rev. Lett.} \textbf{61},
2921-2924 (1988).

\bibitem{Hong-Ou-Mandel}
Hong, C. K., Ou, Z. Y. \& Mandel, L. Measurement of subpicosecond
time intervals between two photons by interference. \textit{Phys.
Rev. Lett.} \textbf{59}, 2044-2046 (1987).

\bibitem{LamasLinares}
Lamas-Linares, A., Howell, J., Bouwmeester, D. Stimulated emission
of polarization-entangled photons. \textit{Nature} \textbf{412},
887-890 (2001).

\end{thebibliography}
\end{document}